\documentclass{aa}
\usepackage{graphics}

\begin{document}

\newcommand{\wat}{H$_2$O}
\newcommand{\esc}{erg s$^{-1}$ cm$^{-2}$}
\newcommand{\escsr}{erg s$^{-1}$ cm$^{-2}$ sr$^{-1}$}
\newcommand{\wc}{W cm$^{-2}$}
\newcommand{\um}{$\mu$m}
\newcommand{\kms}{km s$^{-1}$}
\thesaurus{06(08.06.2;08.09.2 T Tau;08.16.5;09.09.1 T Tau;09.10.1;13.09.4)}
\title{Shock excited far-infrared molecular emission around T~Tau
\thanks{Based on observations with ISO, an ESA project with instruments funded
by ESA Member States 
(especially the PI countries: France, Germany, the Netherlands and the 
United Kingdom) 
with the participation of ISAS and NASA}} 	
\author{L. Spinoglio\inst{1} 
\and T. Giannini\inst{1,2,3} 
\and B. Nisini \inst{2} 
\and M.E. van den Ancker\inst{4} 
\and E. Caux\inst{5}
\and A. M.~Di~Giorgio\inst{1} 
\and D. Lorenzetti \inst{2} 
\and F. Palla \inst{6} 
\and S. Pezzuto \inst{1}
\and P. Saraceno \inst{1} 
\and H.A. Smith \inst{7}
\and G.J. White \inst{8,9}} 
\offprints{ L. Spinoglio, luigi@ifsi.rm.cnr.it}

\institute{Istituto di Fisica dello Spazio Interplanetario, CNR
via Fosso del Cavaliere 100, I-00133 - Roma, Italy
\and 
Osservatorio Astronomico di Roma, via Frascati 33, I-00040 Monte Porzio, Italy
\and
Istituto Astronomico, Universit\`a La Sapienza, via Lancisi 29, I-00161 Roma, Italy
\and
Astronomical Institute ``Anton Pannekoek", University of Amsterdam, Kruislaan
403, NL-1098 SJ Amsterdam, The Netherlands
\and
CESR, BP4346, F-31028 Toulouse Cedex 04, France
\and
Osservatorio Astrofisico di Arcetri, L.go E.Fermi 5, I-50125 Firenze, Italy
\and
Harvard - Smithsonian Center for Astrophysics, 60 Garden Street -
Cambridge, MA, USA
\and
Physics Department, Queen Mary \& Westfield College, University of London, 
Mile End Road - London E1 4NS, UK
\and
Stockholm Observatory, Saltsjobaden, S-133 36, Sweden
}
%
%
\date{Received date ; Accepted date }
%
%
\titlerunning{The shock excited far-IR molecular spectrum of T Tau}
\authorrunning{L. Spinoglio et al.}
\maketitle
\begin{abstract}
The first complete far-infrared spectrum of T Tau has been obtained
with the LWS spectrometer on-board the Infrared Space 
Observatory (ISO), which detected 
strong emission from high-J (J=14-25) CO, para- and ortho-H$_2$O and OH 
transitions over the wavelength range from 40 to 190 \um. 
In addition the [OI]63\um, [OI]145\um~ and  [CII]158\um~ atomic lines were also
detected. 
Most of the observed molecular emission can be explained by a single emission 
region at T$\sim$300-900 K and n$_{\rm H_2}\sim$10$^{5-6}$cm$^{-3}$,
with a diameter of about 2-3 arcsec. This corresponds to a very compact 
region of 300 - 400 AU at the distance of 140 pc. An higher temperature 
component seems to be needed
to explain the highest excitation CO and \wat~ lines.
We derive a water abundance  of 1-7 $\cdot 10^{-5}$ and an OH abundance 
of $\sim$ 3$\cdot 10^{-5}$ with respect to molecular hydrogen,
implying \wat~ and OH enhancements
by more than a factor of 10 with respect to the expected ambient gas abundance.

The observed cooling in the various species amounts to 0.04 L$_{\odot}$,
comparable to the mechanical luminosity of the outflow, indicating that the stellar 
winds could be responsible of the line excitation through shocks.

In order to explain the observed molecular cooling in T Tau in terms of
C-type shock models, we hypothesise that the strong far-ultraviolet
radiation field photodissociates water in favour of OH. 
This would explain the large overabundance of OH observed.

The estimated relatively high density and compactness of the observed
emission suggest that it originates from the shocks taking place
at the base of the molecular outflow emission, in the region where 
the action of the stellar winds  from the two stars of the binary system is important. 

\keywords{Stars: formation; Stars: individual: T Tau; Stars: pre-main sequence; 
ISM: individual objects: T Tau; ISM: jets and outflows; Infrared: ISM: lines and bands}

\end{abstract}


\section{Introduction}

T Tau has been extensively studied from radio to ultraviolet wavelengths
because it has long been considered the prototype of a class of pre-main 
sequence stars. In recent years it has become clear that
T Tau is in reality a very complex system and that it differs from other
T Tauri stars. It is in fact known to be a binary system (Dyck et al. 1982), 
containing  an optical stellar component (T Tau N) and an infrared 
companion 0.7 arcsec to the south (T Tau S), corresponding to 100 AU at 
the 140 pc distance of the Taurus Auriga dark cloud (Wichmann et al. 1998). 
T Tau S dominates the bolometric luminosity of the system (Ghez et al. 1991), 
but it has no detectable optical counterpart to a limiting magnitude of V=19.6, 
suggesting an optical extinction greater than 7 mag (Stapelfeldt et al. 1998). 

Since Herbig's (1950) optical spectra of Burnham's nebula, surrounding 
T Tau several arcsecond across, resembling a so-called HH object, it was clear
that the interaction of stellar winds with the surrounding molecular medium 
is at work. Molecular outflow activity was first mapped by Edwards 
\& Snell (1982) in $^{12}$CO J=1-0 and J=2-1, who found that 95\% of the high 
velocity molecular gas is associated with blueshifted material. The direction 
of the detected outflow is roughly parallel to the line of sight, but 
the emission also 
shows a region extending 2 arcmin to the south and east of T Tau N with a 
secondary peak in the blueshifted wing. 
Higher resolution maps of the $^{12}$CO J=3-2, J=6-5 and C$^{18}$O J=1-0, J=2-1 and
HCO$+$  
emission later showed (Schuster et al. 1993, Momose et al. 1996, Schuster et al. 1997,
Hogerheijde et al. 1998) 
a complex outflow system that could originate from the different components 
of the binary system. 

The fast stellar winds observed through forbidden optical line emission 
(B\"ohm \& Solf 1994), revealed five distinct kinematic components that
suggest that both the primary star and the companion may drive separate bipolar
outflows. A giant Herbig-Haro flow was recently discovered (Reipurth et al. 1997)
around T Tau and is interpreted as originating several thousand years ago from
T Tau S.

Strong and extended ${\rm H_2}$ ro-vibrational emission was found quite early around
T Tau (Beckwith et al. 1978). Recently, high resolution ${\rm H_2}$ imaging 
(Herbst et al. 1996, 1997) indicated that the extended molecular
hydrogen emission arises from the impact on the ambient cloud of two outflow 
systems oriented NW-SE and E-W. These originate from the two stars, each with its
circumstellar disk, and the emission is distributed equally over T Tau N
and T Tau S. 
Infrared adaptive optics observations in ${\rm H_2}$ show instead that
the emission is
concentrated on T Tau S and is interpreted in terms of shocks occurring
as matter accretes onto the circumstellar disk of T Tau S (Quirrenbach \& 
Zinnecker 1997). No firm conclusion is therefore reached on this problem. 

T Tau has associated a substantial amount of mass of dust, it is therefore 
luminous in the millimeter continuum (Adams et al. 1990, Beckwith et al. 1990). 
A circumstellar disk has been detected both with CO interferometry and infrared
scattered light (Weintraub et al. 1989, Momose et al. 1996). Later, millimeter 
continuum interferometry at 0.9 and 3mm (Hogerheijde et al. 1997, Akeson et al. 
1998, respectively) was used to derive a total mass of 0.04 ${\rm M_{\sun}}$ for the 
circumstellar disk around T Tau N and at least 10 times smaller for that 
associated to T Tau S. A circumbinary envelope would also be required to 
fit the continuum energy distribution.

Far-infrared spectroscopy provides powerful diagnostic lines from abundant molecular
species like CO, \wat~ and OH, that can be used to clarify the physical processes
at work in the complex T Tau system. In this 
paper we present the far-infrared spectrum observed from the Long Wavelength
Spectrometer (LWS, Clegg et al. 1996) onboard the Infrared Space Observatory 
(ISO, Kessler et al. 1996). Additional data from the Short Wavelength Spectrometer
(SWS, de Graauw et al. 1996) are also used for discussing the molecular
emission properties. The main results of the SWS are presented
by van den Ancker et al.(1999). The continuum far-infrared spectrum of T Tau
will be discussed in a forthcoming paper (Pezzuto et al. in preparation).                                                                                                                                     
\section{Observations}

\begin{figure}
 \resizebox{\hsize}{!}{\includegraphics{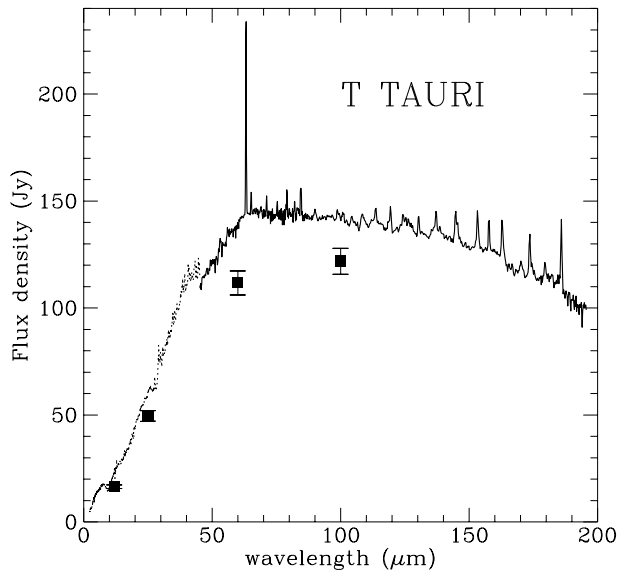}}
\hfill
\caption{The complete ISO LWS and SWS spectrum of T Tau. The IRAS data are also shown for
comparison}
\end{figure}

T Tau 
has been observed with ISO (Infrared Space Observatory) using the LWS  
(Long Wavelength Spectrometer, Clegg et al. 1996).
A full low resolution (R $\sim$ 200) spectrum of the source from 45 to 197 
$\mu$m was obtained during revolution 680, corresponding to September, 25, 1997. 
The beamsize is on average
80 arcsec, depending on the wavelength.
The spectrum was made up of 23 full grating scans oversampled at 1/4 of a 
resolution element, with each 
spectral sample integrated for 11.5 sec, with a total
integration time of 4265 sec. Besides the observation on-source, full grating spectra 
were also collected at four 
off-source positions. In Table 1 we present the journal of the LWS observations,
which includes source and off-source positions and total observing time (OTT).

The raw data were reduced and 
calibrated using version 7 of the LWS pipeline, which achieves an absolute
accuracy of about 30 \% (Swinyard et al. 1998).
Post-pipeline processing was carried out with the ISAP package and included 
removal of spurious signals due to cosmic ray 
impacts and averaging the grating scans of each detector.  

Besides the data measured by the LWS, we also discuss in this paper
the detection of \wat~ and OH emission lines observed by the SWS. The details
of these observations are reported by van den Ancker et al.(1999).

\begin{table}
\caption{Journal of the ISO-LWS observations of T Tau}
\vspace{0.5cm}
\begin{tabular}{lccc}
\hline 
Pos. & 2000.0~~Coordinates & OTT     \\
 & R.A.~~~~~~~~~~~~~~DEC.    &         \\
 & h~~~m~~~s~~~deg~~~'~~~"~~ & (sec)   \\
\hline
T Tau on    & 4:21:59.4  +19:32:06.5  & 4265  \\
T Tau off N & 4:21:59.4  +19:33:46.5  & 1345  \\
T Tau off S & 4:21:59.3  +19:30:26.5  & 1345  \\
T Tau off W & 4:22:06.4  +19:32:06.0  & 1345 \\
T Tau off E & 4:21:52.4  +19:32:07.0  & 1345 \\
\hline
\end{tabular}
\end{table}

\begin{figure*}
 \resizebox{\hsize}{!}{\includegraphics{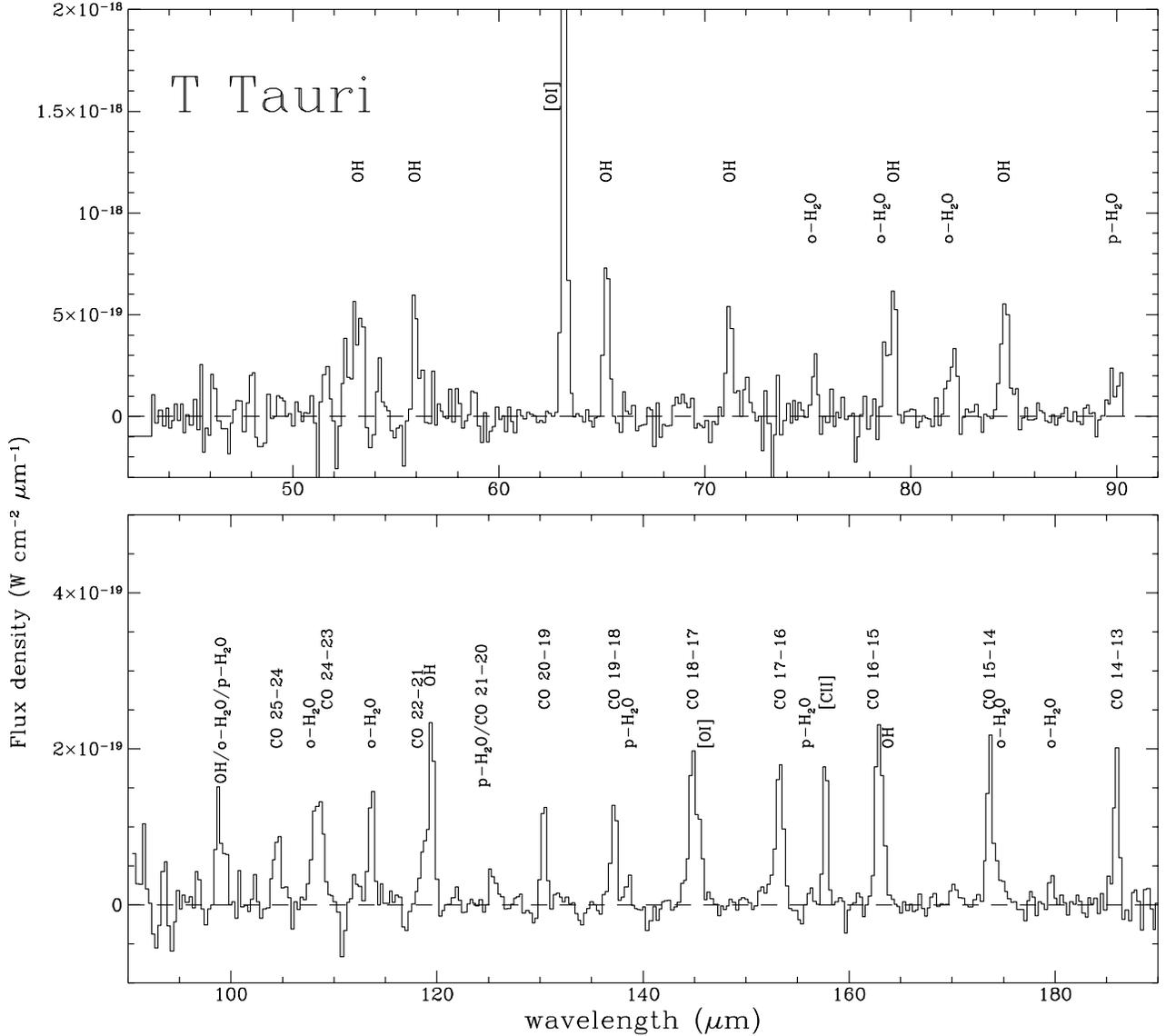}}
\hfill
\caption{The ISO-LWS emission line spectrum of T Tau, from which the continuum 
has been subtracted.}
\end{figure*}

\section{Results}

The 2-200~\um~ far-infrared spectrum of T Tau composed by the ISO LWS and SWS
spectra  is shown in Fig.~1. The displacement (of $\sim$ 15\%) between the 
flux level of the two instruments at $\sim$ 45\um~ is well within the 
calibration uncertainties. In the figure, IRAS photometry is also reported
for comparison. As discussed in van den Ancker et al. (1999), the higher ISO fluxes 
compared to the IRAS data (30-50\%) can be explained by the flare that occurred
to the system in 1990-1991 (Ghez et al. 1991; Kobayashi et al. 1994), after which
the infrared luminosity of T Tau did not return to the pre-outburst value.

The line spectrum (Fig. 2) is very rich in molecular emission lines 
from the rotational spectra of carbon monoxide, water and hydroxyl.
The line fluxes, computed by fitting gaussian profiles to the lines, are listed
in Tables 2-5. 
All of the CO transitions with ${\rm J_{up}}$ = 14-25 appear in the spectrum. 
However, we cannot assign a flux, but only an upper limit, 
to the ${\rm J_{up}}$ = 23 line because it
is blended with the water line o-H$_2$O 4$_{14}$-3$_{03}$.
Water lines are observed both in the ortho (o-) and para (p-) form. Most of the
back-bone lines up to excitation temperatures of more than 700 K are detected
for ortho-\wat~ and up to nearly 400 K for the para-\wat, as well as most of 
the strongest transitions falling in the LWS spectral range.
Also the strong OH lines 
are detected and are particularly strong, with excitation temperatures 
up to 600 K.
[OI]63\um~ and 145\um~ and [CII]158\um~ are the only atomic lines present 
in the ISO-LWS spectrum. 
[OI]63$\mu$m~ emission from T Tau was already detected with the Kuiper Airborne
Observatory from Cohen et al. (1988),
who measured a total flux in a 47$\arcsec$~ aperture of 
(242 $\pm$ 25) $\times$ 10$^{-20}$ W cm$^{-2}$,
which is consistent with our value, which is however much more precise. 
This indicates that the flare occurred in 1990-91 did  not influence the 
far-infrared line emission.
In the four off-source positions only the 
[CII]158\um~ line was detected (see Sect. 3.4 and Table 5).

A large velocity gradient (LVG) code, solving the level population equations 
in a plane parallel geometry (Nisini et al. 1999a, Giannini et al. 1999) 
was used to model
the observed molecular emission line intensities from CO, \wat~ and OH.
As a first approximation, 
the local radiation 
field was not taken into account in the radiation transfer calculations.  
We will however discuss in Sect. 3.3 what is the effect of considering 
the infrared local radiation field in the OH model. 

The model has many free-parameters (gas temperature and density, 
intrinsic width of the line, column density and emitting area or
filling factor, which is related to both number and column densities) that
cannot be easily  constrained simultaneously.
Given the assumption that all of the molecular lines originate from the same
emitting gas, we started our analysis by deriving a range of 
temperature and density that is allowed from fitting the CO lines, that
are likely optically thin.
These parameters are then used to model the water and OH lines, 
which, unlike the CO lines, have high optical depths
due to their strong radiative transitions in the far-infrared.

\subsection{CO emission}

\begin{table} 
\caption{Measured CO line fluxes from the LWS grating spectrum,
with 1$\sigma$ uncertainties. Upper limits are at 3$\sigma$.}
\vspace{0.5cm}
\begin{tabular}{ccccc}
\hline 
 $\lambda _{obs}$ & Line id.& $\lambda _{vac}$  & F & $\Delta$F \\  
 ($\mu$m) &  & ($\mu$m) & \multicolumn{2}{c}{(10$^{-20}$ W cm$^{-2}$)} \\
\hline
       & CO 28-27 & 93.35 & $<$4.1 & \\
       & CO 27-26 & 96.77 & $<$5.3 & \\
       & CO 26-25 & 100.46 & $<$5.3 & \\
104.40 & CO 25-24  & 104.44  & 7.0 & 1.3\\
108.81 & CO 24-23  & 108.76 & 6.7 & 1.1 \\
       & CO 23-22  & 113.46 & $<$10.5$\star$ &  \\
118.58$\dag$ & CO 22-21  & 118.58 & 5.1 & 0.8\\
124.31 & CO 21-20  & 124.19  & 5.2 & 0.9 \\
130.40 & CO 20-19  & 130.37 & 7.7 & 0.5\\
137.17 & CO 19-18  & 137.20 & 9.6 & 0.7\\
144.78$\dag$ & CO 18-17  & 144.78 & 12.6 & 0.5\\
153.24 & CO 17-16  & 153.27 & 14.1 & 0.4\\
162.81$\dag$ & CO 16-15  & 162.81 & 14.0 & 0.7\\
173.63$\dag$ & CO 15-14  & 173.63 & 15.5 & 1.0\\
185.93 & CO 14-13  & 186.00 & 14.3 & 1.0\\
\hline
\end{tabular}
\vspace{0.5cm}

Notes: $\dag$: wavelength was fixed for deblending.\\
$\star$: this line is blended with the o-H$_2$O 4$_{14}$-3$_{03}$
(see text), the total flux has a 1$\sigma$ uncertainty of 
0.5 $\cdot$ 10$^{-20}$ W cm$^{-2}$.
\end{table}

For the CO model, we computed the 
collisional downward rates for levels with ${\rm J_{up}}<$ 60 and 
T $>$ 100 K using the $\gamma_{J0}$ coefficients taken 
from McKee et al. (1982), while the upward rates were computed using 
the principle of detailed balance. Radiative decay rates were 
taken from Chackerian \& Tipping (1983).

The distribution of the observed CO line fluxes as a function
of the rotational quantum number is shown in Fig. 3.
Because the CO lines are optically thin, their emission, in the
LVG model considered, does not
depend on the velocity gradient and thus on the assumed line-width.
We have considered for our fit only the transitions
with ${\rm J_{up}}$ less than 22. 
Our data are consistent with gas temperatures ranging from T = 300 to 900 K 
and molecular hydrogen densities of ${\rm n = 10^{5-6} cm^{-3}}$. 
The two  extreme models consistent with the data have:
\begin{itemize}
\item [] T = 300~K and ${\rm n_{H_{2}} = 4 \cdot 10^{6} cm^{-3}}$;
\item [] T = 900~K and ${\rm n_{H_{2}} = 2 \cdot 10^{5} cm^{-3}}$.
\end{itemize}

Fig. 3 shows that the transitions
with ${\rm J_{up}}$ = 24 and 25 have a flux level which is too high to be 
explained by the same gas component of the other lines and may indicate
the presence of a warmer gas emission. This warmer component, which cannot easily 
be constrained  by the higher ${\rm J_{up}}$ transitions observed, could also affect the 
${\rm J_{up}}$ = 21 and 22 lines. 
However fitting the component
covering the lines 14 $\le {\rm J_{up}}\le$ 20, results in the same  parameters
as the low temperature model above.

\begin{figure}
 \resizebox{\hsize}{!}{\includegraphics{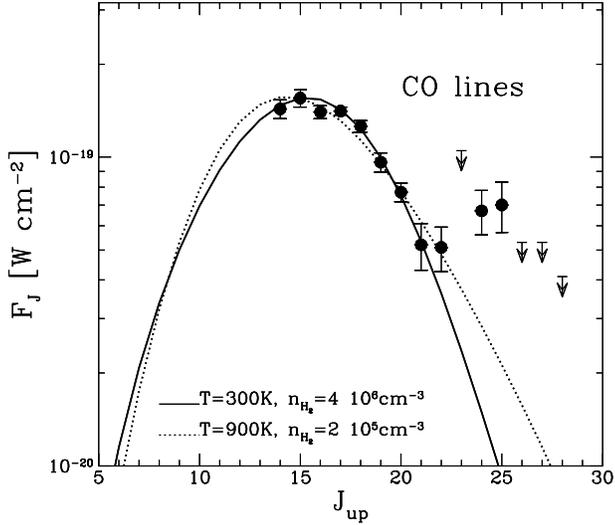}}
\caption{Model fits through the observed CO lines. The range of
temperatures and densities compatible with the observations are indicated.
The higher observed J lines (J=24, 25 and 26) have fluxes too high to be
fitted by the same parameters as the other lines, suggesting the 
presence of a second component.}
\end{figure}

\begin{figure}
 \resizebox{\hsize}{!}{\includegraphics{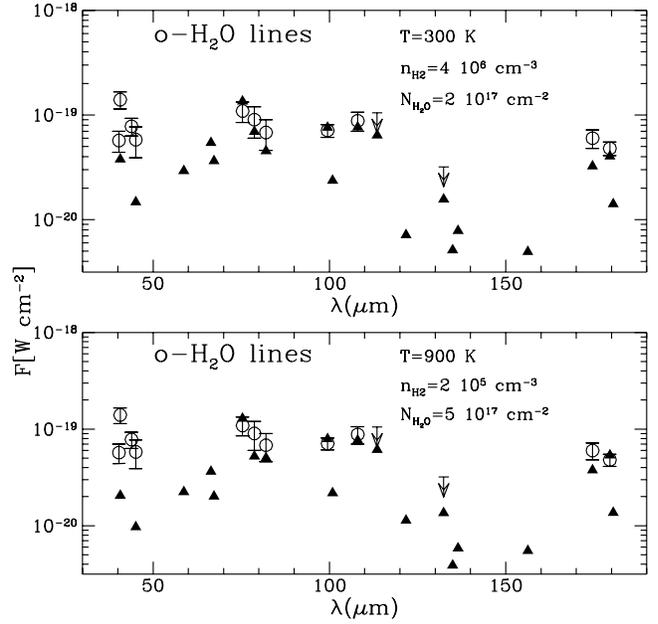}}
\caption{Comparison of the modeled ortho-\wat~ line fluxes (filled triangles) 
with those observed (open circles) for the two models considered.}
\end{figure}

\begin{figure}
 \resizebox{\hsize}{!}{\includegraphics{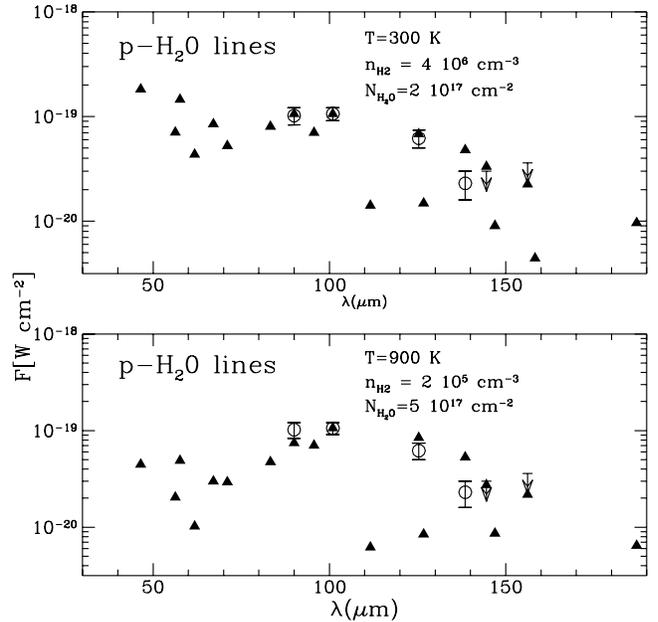}}
\caption{Comparison of the modeled para-\wat~ line fluxes (filled triangles) 
with those observed (open circles) for the two models considered.}
\end{figure}

\subsection{\wat~ emission}

\begin{table}
\caption{Measured water line fluxes from the SWS (upper list) and 
LWS (lower list) grating spectrum, with 1$\sigma$ uncertainties. 
Upper limits are at 3$\sigma$.}
\vspace{0.5cm}
\begin{tabular}{ccccc}
\hline 
 $\lambda _{obs}$ & Line id.& $\lambda _{vac}$  & F & $\Delta$F \\  
 ($\mu$m) &  & ($\mu$m) & \multicolumn{2}{c}{(10$^{-20}$ W cm$^{-2}$)} \\
\hline
25.940 & o-H$_2$O 5$_{41}$-4$_{14}$ & 25.940 & 2.4 & 0.6\\
29.838 & o-H$_2$O 7$_{25}$-6$_{16}$ & 29.836 & 4.9 & 1.1\\
       & o-H$_2$O 4$_{41}$-3$_{12}$ & 31.771 & $<$3.0 &  \\
40.342 & o-H$_2$O 6$_{43}$-5$_{32}$ & 40.337 & 5.7 & 1.3\\
40.688 & o-H$_2$O 4$_{32}$-3$_{03}$ & 40.688 & 14.1 & 2.6\\
43.894$\dag$ & o-H$_2$O 5$_{41}$-4$_{32}$ & 43.894 & 7.8 & 1.5\\
45.116 & o-H$_2$O 4$_{32}$-3$_{03}$ & 45.111 & 5.8 & 1.9 \\
\hline
75.35  & o-H$_2$O 3$_{21}$-2$_{12}$ & 75.38 & 10.9 & 2.4 \\%
78.74$\dag$ & o-H$_2$O 4$_{23}$-3$_{12}$ & 78.74 & 9.0 & 3.0 \\%
82.03$\dag$  & o-H$_2$O 6$_{16}$-5$_{05}$ & 82.03 & 6.8 & 2.2 \\%
90.05  & p-H$_2$O 3$_{22}$-2$_{11}$ & 89.99 & 10.2 & 1.9\\
99.46$\dag$  & o-H$_2$O 5$_{05}$-4$_{14}$ & 99.49 & 7.1 & 1.0\\%
100.92 & o-H$_2$O 5$_{14}$-4$_{23}$ & 100.91 & 10.6 & 1.5\\%
       & p-H$_2$O 2$_{20}$-1$_{11}$ & 100.98 & & \\%
108.11 & o-H$_2$O 2$_{21}$-1$_{10}$ & 108.07 & 8.8 & 1.8 \\%
113.65 & o-H$_2$O 4$_{14}$-3$_{03}$ & 113.54 & 10.5$\star$ & \\%
125.39 & p-H$_2$O 4$_{04}$-3$_{13}$ & 125.35 & 6.2 & 1.2 \\
       & o-H$_2$O 4$_{23}$-4$_{14}$ & 132.41 & $<$3.2 & \\%
138.62 & p-H$_2$O 3$_{13}$-2$_{02}$ & 138.53 & 2.3& 0.7\\%
       & p-H$_2$O 4$_{13}$-3$_{22}$ & 144.52 & $<$3.0 & \\%
       & p-H$_2$O 3$_{22}$-3$_{13}$ & 156.19 & $<$3.6& \\%
174.62$\dag$ & o-H$_2$O 3$_{03}$-2$_{12}$ & 174.63 & 6.0 & 1.2\\%
179.58 & o-H$_2$O 2$_{12}$-1$_{01}$ & 179.53 & 4.8 & 0.7\\%
\hline
\end{tabular}
\vspace{0.5cm}

Notes: $\dag$: wavelength was fixed for deblending.\\
$\star$: this line is blended with the CO 23-22 line
(see text), the total flux has a 1$\sigma$ uncertainty of 
0.5 $\cdot$ 10$^{-20}$ W cm$^{-2}$.
\end{table}

As outlined in the previous Sect., we adopted the
temperature and density as derived from the CO lines
models to fit the observed \wat~ line fluxes.

We considered in the computation 45 levels for both the ortho and para species
({\it i.e.} excitation energies up to $\sim$ 2000 K):
radiative rates are taken from Chandra et~al. (1984) while the H$_2$O-H$_2$
collision rates are derived from Green et~al. (1993). We assumed
an ortho/para abundance ratio of 3, equal to the ratio of the statistical
weigths of their nuclear spins.

The other parameters that enter in the model are 
the velocity gradient in the region (dV/dr) 
and the projected area of the emission region. 
The optical depth in the lines is directly proportional to the water column
density (N(\wat)). 
Since the ratios of different lines depend on their relative optical depths, 
we can use them to constrain dV/N(\wat). On the other hand, the
absolute line intensity depends on both the column density and the projected area 
of the emission region, 
and therefore if we assume a velocity linewidth dV, we can estimate
both the column density and the emission region size. 

The results of the model fitting are shown in Fig. 4 and Fig. 5 for ortho-
and para-\wat~ respectively. 
Almost all the lines in the LWS wavelength interval are well reproduced by the
model. The differences in the flux level predicted in the two extreme models
are quite small, indicating that the \wat~ emission is not very sensitive to 
the exact value of temperature and density in the range. On the other hand, 
we note that the ortho-\wat~ lines shortward
of  50\um~ are not fitted by our models. These lines, originated 
by levels at energies higher than 500 K, are brighter than
our predictions, indicating that a warmer component might be required,
as also suggested from the higher J CO lines.

An estimate of the intrinsic linewidth dV can be given if we relate the 
observed emission with the outflow/wind activity taking place in the close 
environment of the T Tau binary system. 
The molecular outflow has been traced by different lines at near
infrared and millimeter wavelengths (${\rm H_2, CO, HCO^{+}}$), showing 
linewidths of a few \kms. 
Adopting a velocity of 10~\kms, close to the outflow velocities of 7.9~\kms (red lobe)
and 9.7~\kms (blue lobe) measured by Levrault (1988) and those measured by 
Hogerheijde et al. (1998) in ${\rm ^{13}CO}$ 3-2 (of 12. and 15.6~\kms for the
red and blue lobes , respectively), we derive a water 
column density of (2-5)10$^{17}$ cm$^{-2}$, while the projected area
is (4-9) arcsec$^2$. This corresponds to a diameter of only 300 - 400 AU, 
assuming spherical simmetry. The compactness of this emission region will 
enable us to put constraints on the physical mechanisms responsible of the 
observed emission (see Sect. 4).

Using this emission area, 
the CO column density that we derive from the observed CO absolute
line fluxes is N(CO) = (0.7-2.0)10$^{18}$ cm$^{-2}$ and therefore an \wat/CO 
abundance ratio of $\sim$ 0.1 - 0.7. 
Assuming a standard CO abundance of 10$^{-4}$, the water abundance with 
respect to H$_2$ is $\sim$ (1 - 7) $\cdot$ 10$^{-5}$. 
This value implies an enhancement with respect 
to the expected abundance in the ambient gas of at least a factor of 10 
(e.g. Bergin et al. 1998).\\

High \wat~ abundances are common in young stellar objects: 
the ISO spectrometers have in fact found strong emission from 
gas-phase \wat~ from massive young stars (Harwit et al. 1998, Gonz\'alez-Alfonso et
al. 1998) and from low mass outflow driving sources 
(Liseau et~al. 1996, Saraceno et~al. in preparation, Ceccarelli et~al. 1998) 
with abundances in the range 1-5$\cdot$10$^{-5}$, rising to values as high 
as $\sim$ 5$\cdot$10$^{-4}$ in L1448mm (Nisini et~al. 1999b) and Orion 
(Harwit et~al. 1998).

\subsection{OH emission}

\begin{table*}
\caption{Measured OH line fluxes from the SWS (upper list) and 
LWS (lower list) grating spectrum, with 1$\sigma$ uncertainties. Upper limits are at 3$\sigma$.}
\vspace{0.5cm}
\begin{tabular}{cccccc}
\hline 
 $\lambda _{obs}$ & FWHM & Line id.& $\lambda _{vac}$  & F & $\Delta$F \\  
 ($\mu$m) & ($\mu$m) & & ($\mu$m) & \multicolumn{2}{c}{(10$^{-20}$ W cm$^{-2}$)} \\
\hline
28.931& 0.020 &  OH $^2\Pi _{1/2}$7/2-$^2\Pi _{3/2}$5/2 & 28.939 & 2.7 &0.9\\
      &       &  OH $^2\Pi _{1/2}$5/2-$^2\Pi _{3/2}$3/2 & 34.603/34.629 & $<$6.4 & \\
43.950$\dag$& 0.045$\dag$  & OH $^2\Pi _{1/2}$7/2-$^2\Pi _{3/2}$7/2 & 43.950 & 5.2 & 1.5 \\
\hline
53.17& 0.75 & OH $^2\Pi _{3/2}$11/2-$^2\Pi _{3/2}$9/2 & 52.93/53.06&61.2& 4.4\\
     &  & OH $^2\Pi _{1/2}$3/2-$^2\Pi _{1/2}$3/2 & 53.26/53.35 &    & \\
55.91$\dag$& 0.35 & OH $^2\Pi _{1/2}$9/2-$^2\Pi _{1/2}$7/2 & 55.89/55.95& 17.9 & 3.3\\
65.23& 0.29 & OH $^2\Pi _{3/2}$9/2-$^2\Pi _{3/2}$7/2 & 65.13/65.28 & 23.7 & 1.0 \\
71.22& 0.30 & OH $^2\Pi _{1/2}$7/2-$^2\Pi _{1/2}$5/2 & 71.17/71.22 & 18.2 & 3.3 \\
79.16$\dag$& 0.29 & OH $^2\Pi _{1/2}$1/2-$^2\Pi _{1/2}$3/2 & 79.12/79.18 & 17.8 & 3.0 \\
84.59& 0.59 & OH $^2\Pi _{3/2}$7/2-$^2\Pi _{3/2}$5/2 & 84.42/84.60 & 25.9 & 2.4 \\
98.73$\dag$& 0.6 & OH $^2\Pi _{1/2}$5/2-$^2\Pi _{1/2}$3/2 & 98.73 & 9.8 & 1.0\\
119.44& & OH $^2\Pi _{3/2}$5/2-$^2\Pi _{3/2}$3/2 & 119.23/119.44 & 14.9 & 1.5\\
163.12$\dag$&0.6 & OH $^2\Pi _{1/2}$3/2-$^2\Pi _{1/2}$-1/2 & 163.12/163.40 & 8.9 & 0.7 \\
\hline
\end{tabular}
\vspace{0.5cm}

Notes: $\dag$: wavelength was fixed for deblending.
\end{table*}

\begin{figure}
 \resizebox{\hsize}{!}{\includegraphics{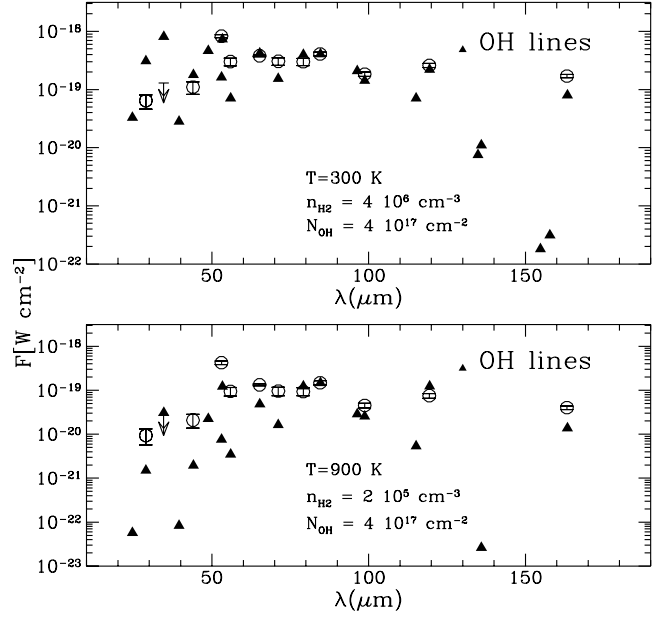}}
\caption{Comparison of the modeled OH line fluxes (filled triangles) 
with those observed (open circles).}
\end{figure}

For the OH models, we have considered 20 levels. The collisional downward 
rates are from Offer \& van Dishoeck (1992) and the radiative decay rates 
are from the HITRAN catalogue (Rothman et~al. 1987).

If we adopt the same parameters as derived from the above analysis
also for the OH, we find that a better agreement between data and models is
achieved with the lower temperature model (T = 300 K).
The estimated OH column density is 
N(OH)  $\sim$ $4\cdot 10^{17}$ cm$^{-2}$ and therefore a 
 X(OH) $\sim$ $2.7\cdot 10^{-5}$. 
The results of the OH model fitting are shown in Fig. 6. As can be seen
from this figure, not all the lines can be reproduced by the models, in
particular both the 163\um~ line and those shortward of 60\um.

Because T Tau is  
relatively bright in the continuum at the far infrared wavelengths (see Fig.1), 
such a discrepancy could be due to the pumping from the local thermal radiation
field. To test this possibility 
we also computed models for the OH transitions including the field originated
from dust at a temperature of 300K, using the model of Cesaroni \& Wamsley (1991).
 We found that the inclusion of the local infrared radiation 
field increases the emission in the lines with $\lambda$ less than 100\um~ and
the 163\um~line. 

\begin{table*}
\caption{Measured atomic line fluxes from the LWS grating spectrum with 
uncertainties.}
\vspace{0.5cm}
\begin{tabular}{cccccc}
\hline 
 Pos. & $\lambda _{obs}$ & Line id.& $\lambda _{vac}$  & F & $\Delta$F \\  
  & ($\mu$m) &  & ($\mu$m) & \multicolumn{2}{c}{(10$^{-20}$ W cm$^{-2}$)} \\
\hline
on&63.21  & [OI] $^3P_1\rightarrow {^3P_2}$ & 63.18  & 230.6 & 1.0\\
on&145.52$\dag$ & [OI] $^3P_0\rightarrow {^3P_1}$ & 145.52 & 8.15 & 0.78 \\
on&157.76 & [CII] $^2P_{3/2}\rightarrow {^2P_{1/2}}$  & 157.74 & 10.6 & 0.5\\
off-N& 157.80& [CII] $^2P_{3/2}\rightarrow {^2P_{1/2}}$  & 157.74 & 7.9&0.7\\
off-S& 157.79& [CII] $^2P_{3/2}\rightarrow {^2P_{1/2}}$  & 157.74 & 8.8 & 1.2\\
off-W& 157.76& [CII] $^2P_{3/2}\rightarrow {^2P_{1/2}}$  & 157.74 & 8.8 & 0.9\\
off-E& 157.74& [CII] $^2P_{3/2}\rightarrow {^2P_{1/2}}$  & 157.74 & 9.1 & 0.7\\
\hline
\end{tabular}
\vspace{0.5cm}

Notes: $\dag$: wavelength was fixed for deblending.
\end{table*}

\subsection{Atomic emission}

The detection of [CII]158\um~ in the four off-source positions around T Tau
(see Table 5) clearly shows that most of the ionized carbon emission 
($\sim 80\%$) is from
an extended region and not originated in the vicinity of T Tau.
The intrinsic emission in the LWS beam centered on T Tau is about 
2 $\cdot$ 10$^{-20}$ W cm$^{-2}$. This implies a ratio [OI]63\um/[CII]158\um~
of 115, greatly in excess of that expected from photodissociation region
models (Kaufman et~al.1999; Burton et~al. 1990).


Finally the ratio of [OI]63$\mu$m/145$\mu$m=28.3 is such that no oxygen self-absorption
should occur, as it often appears to be the case towards pre-main sequence sources 
(Saraceno et al. 1998). This suggests that
there is no cold gas in front of the 
source, in accordance with the geometry of the outflow directed towards 
the observer.

As outlined in van den Ancker et al. (1999), based on a larger set of 
fine-structure
lines detected at shorter wavelengths, we argue that the atomic line emission 
observed is consistent with the presence of J-type dissociative shocks.
On the other hand, the excess [CII]158\um~ emission on-source could also be
due to a local photodissociation region (PDR), possibly originated from the
far-UV field of T Tau (see Sect. 4).

\subsection{Total cooling}

Table 6 summarises the physical quantities derived from the observed 
molecular spectra of CO, \wat~ and OH, adopting the two models considered: 
the column densities and the total cooling luminosities are given for each
molecular species. 
We also give the observed values of the cooling
derived from the sum of all the detected fluxes. 
For deriving L$_{\rm H_2}$, we used the line fluxes reported in van den 
Ancker et al. (1999) and the line ${\rm H_2}$ 1-0 S(1) at 2.12\um, given in Carr (1990).
A comparison between the observed and modeled cooling shows that the
observations of CO, \wat~ and OH can account for most of the modeled cooling.
The underestimate of the OH cooling by the 900 K model confirms that this latter
is probably inadeguate to explain the observed OH emission. 
The total radiated cooling observed from these species, including [OI],
sums up to about 0.04 L$_{\odot}$, and has to be considered as a lower limit. 
Taking the
outflow parameters from the literature (Levreault 1988; 
Mundt 1984) we derive a total mechanical luminosity of about 
0.05 L$_{\odot}$ (we have taken an average velocity of 10 ${\rm km s^{-1}}$, 
a total outflow mass of  0.22 M$_{\odot}$, and a dynamical timescale of 40,000
years). 
The radiative luminosity observed in  the far-infrared is therefore comparable
to the outflow mechanical luminosity. This is expected if the stellar winds from the 
stars are driving the outflows and the shocks, traced by the far-infrared 
lines, accelerate the ambient medium into the molecular outflow 
(Davis \& Eisl\"{o}ffel 1995).

\begin{table}[htb]
  \caption{Physical parameters of the molecular and atomic emission}
  \label{tab:table}
  \begin{center}
    \leavevmode
    \footnotesize
    \begin{tabular}[h]{lccc}
      \hline \\
                                    & observed$\star$    & ``lower T''  &  ``higher T''     \\
      \hline \\
      Temperature T(K)       &  & 300     & 900  \\
      Density ${\rm n_{H_2} (cm^{-3})}$ & & 4$\cdot$10$^{6}$ & 2$\cdot$10$^{5}$ \\
      N$_{\rm CO}$ (cm$^{-2}$)    &  & $2\cdot 10^{18}$ & $7\cdot 10^{17}$\\
      N$_{\rm H_2O}$ (cm$^{-2}$)  &  & $2\cdot 10^{17}$ & $5\cdot 10^{17}$ \\
      N$_{\rm OH}$ (cm$^{-2}$)    &   & 4$\cdot$10$^{17}$  & $4\cdot 10^{17}$ \\
      L$_{\rm CO}$ (L$_{\odot}$)  & 0.0067         & 0.0096 & 0.010\\
      L$_{\rm H_2O}$ (L$_{\odot}$)&  0.008        & 0.014  & 0.009 \\
      L$_{\rm OH}$ (L$_{\odot}$)  &     0.012     & 0.019  & 0.009 \\
      L$_{\rm [OI]}$ (L$_{\odot}$) &   0.014    & ---  & --- \\
      L$_{\rm H_2}$ (L$_{\odot}$)  &    0.005     & ---  & --- \\

      \hline \\   
\end{tabular}
\vspace{0.5cm}

Notes: $\star$: The observed cooling is computed by summing all the 
detected lines.

\end{center}
\end{table}

\section{Discussion}

\begin{figure}
\resizebox{\hsize}{!}{\includegraphics{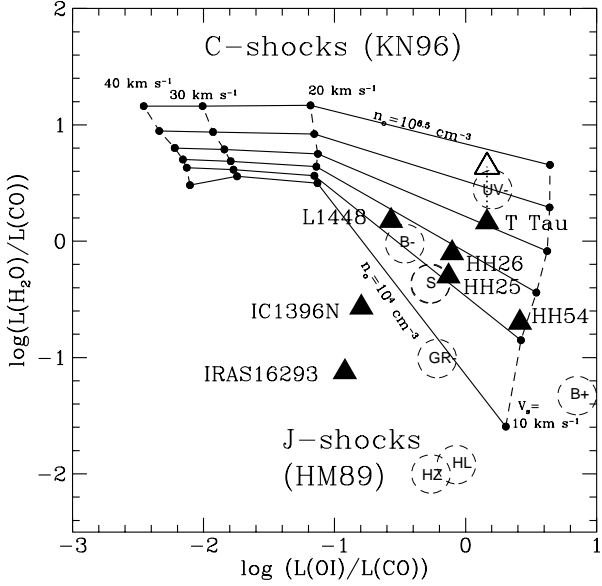}}
\caption{Total cooling from water lines versus [OI]63\um~ cooling,
both normalized to the CO (high-J) cooling in few objects, including T Tau 
(filled triangles) and according to shock models.
The open triangle shows the position that T Tau would have if the water cooling 
were increased by a factor 3.
C-type shock models of 
Kaufman \& Neufeld (1996) are shown in a grid where shock velocity increases 
from the right to the left 
from ${\rm 10 < v_s < 40 ~km~s^{-1}}$ (dashed lines) and
density from the bottom to the top from 10$^4$ to 10$^{6.5}$ ${\rm cm^{-3}}$ 
(solid lines). 
J-type shock models (Hollenbach \& McKee 1989) are shown for comparison as 
large dashed circles, where S is the standard model, with preshock density of
${\rm n_o}$ = 10$^{5}$ ${\rm cm^{-3}}$, shock velocity of 
${\rm v_s = 80 ~km~s^{-1}}$ and magnetic field of ${\rm B = 158{\mu}G}$; 
UV-: the far-ultraviolet field is reduced by a factor of 10;
B-: the magnetic field is reduced by a factor 10; 
B+: the magnetic field is increased by a factor 10; 
GR-: the grain size distribution is extended down to 10 $\AA$; 
HL: ${\rm H_2}$ formation on grains is equal to zero;
HZ: ${\rm H_2}$ formation heating is set to zero.}
\end{figure}

\begin{figure}
 \resizebox{\hsize}{!}{\includegraphics{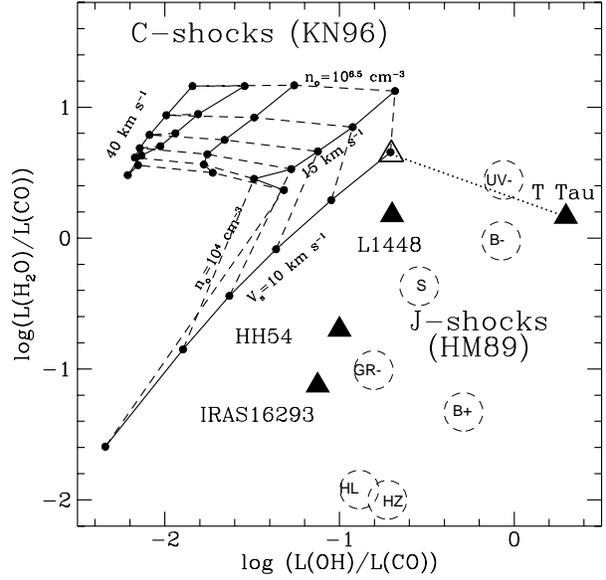}}
\caption{Total cooling from water lines versus total cooling from OH lines,
both normalized to the CO (high-J) cooling in few objects, including T Tau 
(filled triangles) and according to shock models. 
The open triangle shows the position that T Tau would have if the water cooling 
were increased by a factor 3 and the OH cooling decreased by a factor 10.
C-type shock models of 
Kaufman \& Neufeld (1996) are shown in a grid as a function of density and 
shock velocity. J-type shock models 
(Hollenbach \& McKee 1989) are shown as large dashed circles 
(see caption of Fig.7).}
\end{figure}

\begin{figure}
 \resizebox{\hsize}{!}{\includegraphics{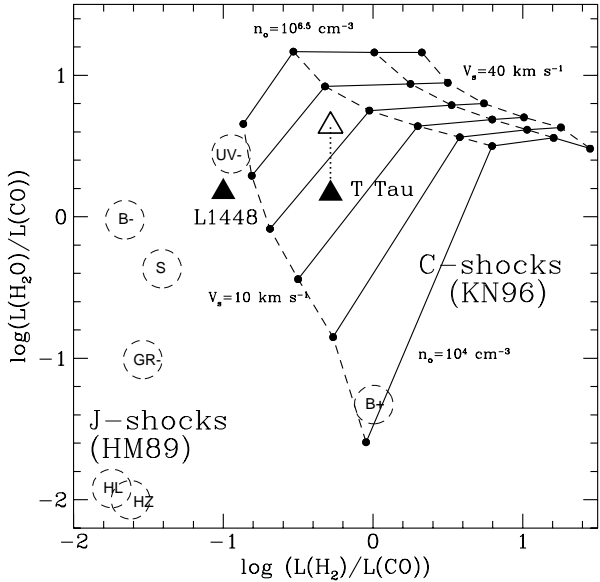}}
\caption{Total cooling from water lines versus total cooling from ${\rm H_2}$
lines, both normalized to the CO (high-J) cooling in T Tau and L1448 (Nisini 
et al. 1999b)(filled triangles) and according to shock models. 
The open triangle shows the position that T Tau would have if the water cooling 
were increased by a factor 3.
C-type shock models of 
Kaufman \& Neufeld (1996) are shown in a grid where density increases upward and 
shock velocity from the left to the right. J-type shock models 
(Hollenbach \& McKee 1989) are shown as large dashed circles (see caption of Fig.7).}
\end{figure}

Once the physical conditions
of the emitting gas in the vicinity of T Tau have been established,
we can now proceed to compare the results with existing models of shock excitation.
The far-infrared line emission of young stellar objects is mainly originated from two  
physical processes: the excitation  from photoionized and photodissociated (PDR) 
regions (Tielens \& Hollenbach 1985) and the shock 
excitation produced by the 
interaction  of supersonic winds with the ambient medium.
Depending on wind velocity, magnetic field and ion density, 
two kinds of shocks with different far-infrared spectra are predicted from models:
\begin{itemize}
\item[i)] high velocity dissociative J shocks (e.g. Hollenbach \& McKee 1989), 
in which  temperature, density and velocity have a discontinuous jump 
(J) on the  shock front, molecules are dissociated 
and atomic lines are the dominant coolants;  
\item[ii)] low velocity non-dissociative C shocks
 (e.g. Kaufman  \& Neufeld  1996, Draine et al. 1983) in which 
the ion Alfv\'en velocity is larger than the shock velocity and 
the magnetic field transmits energy faster than the shock 
velocity; in this case temperature, density and velocity have a 
continuous (C) variation and molecules are the dominant coolants.
\end{itemize}

Instead of using  intensities of many tens of different lines, we can
obtain a better comparison between our data and the shock model predictions 
using the total cooling from a single species. 

In Fig.7 we show the water cooling as a function of the  [OI]63\um~ cooling, 
both normalized to the high-J CO cooling. 
The C-type shock models of Kaufman \& Neufeld (1996)
are considered. The J-type shock models (Hollenbach \& McKee 1989) are also indicated
(see figure caption for details). Together with the position of T Tau, we also
show in this figure the positions of L1448 (Nisini et al. 1999b), IC1396 (Saraceno
et al. 1999), IRAS16293-2422 (Ceccarelli et al. 1998), and the Herbig-Haro objects
HH54 (Liseau et al. 1996), HH25 and HH26 (Benedettini et al. 1998).

T Tau is in a central position, showing that both C-type and
J-type shock models could explain the observations. It has to be noted, 
however, that according to Kaufman \& Neufeld models, its position implies a 
shock velocity between 10 and  ${\rm 15 ~km~s^{-1}}$, in a range where
water production is triggered but it is not at its maximum efficiency. 
We can see in the figure that other pre-main sequence sources also cluster
in the same region of T Tau, indicating that these shock conditions are
fairly common in the environment of young stellar objects (Nisini et al. 1998).

Fig. 8 shows the water cooling 
as a function of the OH cooling , both normalized to the high-J
CO cooling. As before, we consider both J-type and C-type shocks.
The position of T Tau in this plot appears to be consistent with J-type shocks.
The main reason why C-shocks fail to reproduce the observed cooling is
the overabundance of OH molecules, by at least a factor 10. Only a model 
which includes the effects of the presence of a high UV field from T Tau itself
can reproduce the observed values.

Fig. 9 shows the water cooling 
as a function of the  ${\rm H_2}$ cooling, both normalized to the high-J
CO cooling. T Tau lies exactly in the center of the region of 
C-type shocks and cannot be explained by J-type shocks. 
As expected, the strong {\rm H$_2$} emission cannot be accounted for by 
J-type shocks. We therefore rule out the hypothesis that the major responsible 
of the observed excitation are J-shocks.

Because photodissociation of water by an UV field, that is not included in 
shock models, can convert water to OH,
we suppose that the overabundance of the OH molecule is due only
to the strong far-UV radiation field associated to T Tau (Herbig \& Goodrich, 1986).
The photodissociation cross Sect. of water 
at the $Ly\alpha$ frequency is in fact ten times larger than the one of OH
(van Dishoeck \& Dalgarno 1984). 
A similar situation has been found in supernova remnants, where OH 1720 MHz
emission is explained as originated from C-type shocks, allowing that the
action of an UV field creates sufficient OH from water dissociation
(Wardle et al. 1998; Lockett et al. 1999).

If this is the case, the far-infrared molecular emission 
spectrum of T Tau is primarily due to C-type shocks. 
To reconcile the OH observations with C-type shock models we need that
the OH abundance, and thus its total cooling, were reduced by a factor 10
in favor of water cooling. 
From Table 6, if we reduce by a factor 10 the
OH column density and we increase of the corresponding amount that one of 
the water, 
passing from $2\cdot 10^{17}$ to $5.6\cdot 10^{17}$, we will increase the
total water cooling by a factor of about 3. 

Increasing the water cooling by this factor and decreasing the OH cooling
by a factor 10 would move the position of T Tau in the three diagrams of 
Fig.7, 8 and 9 in positions fully consistent with C-type shock models (see
the open triangles in the figures), at pre-shock 
densities of about 10$^{5.5-6}$ ${\rm cm^{-3}}$
and shock velocities of ${\rm 10 < v_s < 20 ~km~s^{-1}}$. Comparing the
pre-shock densities of these models with the densities derived from our LVG models,  
we found that
a moderate compression factor ($\le$ 10) would be required. In support
of the fact that the molecular emission is due to C-type shocks 
is the evidence of the presence of strong magnetic fields in the outflow
region associated with T Tau S (Ray et al. 1997). 

Our conclusion that the observed far-infrared molecular emission from T Tau can 
be explained by C-type shocks and that the atomic emission is probably
originated in J-type shocks is in agreement with the findings of van den Ancker
et al. (1999). Their models, based on the near to mid-infrared ${\rm H_2}$ 
emission imply two temperature components at 440 K and 1500 K, which again are
roughly in agreement with our finding of two components: one ranging from
300 K to 900 K and another at an higher temperature traced by the higher 
${\rm J_{up}}$ CO transitions.

As to the origin of the C-type shocks,
responsible for the observed far-infrared molecular emission, 
we know from the LVG models that the emission region has a size of only few
hundreds AU, assuming a spherical geometry.
This implies that the shocks occur in a very compact region, presumably
very close to the binary system. There are at least three mechanisms
not mutually exclusive to explain the origin of the shocks: 
\begin{enumerate}
\item from the interaction region of winds coming from the two stars;
\item from disk accretion on the youngest component of the binary system;
\item from the interaction of the stellar wind with the molecular material 
in the circumstellar envelope.
\end{enumerate}
The first possibility seems the best one for originating fast (${\rm v_s \sim 50~km~s^{-1}}$)
dissociative J-type shocks, because the wind interaction would occur close
to the stars where velocities are supposed to be high and the stellar field 
strong to dissociate molecules.
The second possibility has already been suggested by Quirrenbach \& Zinnecker (1997)
to explain the near-infrared ${\rm H_2}$ extended emission.
The third one is probably at work in any case, because it is needed to explain
the strong emission from CO, \wat, OH, as well as that from ${\rm H_2}$ (van den 
Ancker et al. 1999).

\section{Conclusions}

To summarise our results, we list the main findings of this study:
\begin{enumerate}

\item The far-infrared spectrum associated to the binary system of T Tau shows
strong emission from CO, \wat~ and OH molecules.

\item Optically thin CO emission lines from high-J transitions are used to constrain the
physical regimes of the gas: T$\sim$300-900 K and n$_{\rm H_2}\sim$10$^{5-6}$cm$^{-3}$.
The detection of CO lines with ${\rm J_{up}}$ of 24 and 25 seems to indicate that
a warmer component is also needed.

\item \wat~ and OH emission is consistent with such conditions, however the higher 
excitation lines at the shorter wavelenghs are not well fitted by these models,
indicating that higher temperature gas should also be present, in agreement with
the CO emission.

\item From the assumption that all the far-infrared molecular emission observed
originate from the same region a very compact size of 300 - 400 AU of diameter
is derived.
 
\item The detection of [CII]158\um~ off-source at large distances from T Tau
shows that most of the ionised carbon emission ($\sim 80\%$) is from
an extended region and not originated in the vicinity of T Tau.
The intrinsic emission in the LWS beam centered on T Tau 
implies a ratio [OI]63\um/[CII]158\um~
of 115, much in excess of what is expected from photodissociation region
models. This emission is probably due to J-type shocks, however it is not 
ruled out the possibility of a contamination by a PDR.

\item The total cooling observed in the various species is
comparable to the mechanical luminosity of the outflow, indicating that the stellar 
winds could be the ultimate responsible of the line excitation through shocks.

\item The study of the cooling from the main radiative components ([OI], CO, \wat,
OH and ${\rm H_2}$) is used to show that our data are consistent with C-type
shock models, only if we assume an overabundance of the OH molecule
(by a factor 10) probably due to the intense far-ultraviolet radiation field.
 
\item We finally consider the origin of the shocks responsible for the
far-infrared emission: colliding winds from the components of the binary 
system, disk accretion in the more embedded star and wind interaction 
with the molecular ambient cloud. Due to the large beam size of the LWS
instrument, we are unable to specify which of the processes is the 
dominant one.

\end{enumerate} 

\begin{acknowledgements}
We wish to thank the LWS Consortium and the ISO staff at VILSPA (ESA, 
Villafranca, Spain) for having operated the LWS instrument and the ISO
satellite. 
We thank Riccardo Cesaroni for running his OH code to evaluate the effects
of a local radiation field. MvdA acknowledges financial support 
from NWO grant 614.41.003 and through a NWO {\em Pionier} grant to L.B.F.M. 
Waters.
\end{acknowledgements}

\end{document}